# On Hallucinations in Artificial Intelligence–Generated Content for Nuclear Medicine Imaging (the DREAM Report)

Menghua Xia[1,2], Reimund Bayerlein[3,4], Yanis Chemli[1,2], Xiaofeng Liu[1,2], Jinsong Ouyang[1,2], MingDe Lin[1,5], Georges El Fakhri[1,2], Ramsey D. Badawi[3,4], Quanzheng Li[6], and Chi Liu[1,2]

*[1]Department of Radiology and Biomedical Imaging, Yale University School of Medicine, New Haven, Connecticut; [2]Yale Biomedical Imaging Institute, Yale University, New Haven, Connecticut; [3]Department of Biomedical Engineering, University of California Davis, Sacramento, California; [4]Department of Radiology, University of California Davis, Sacramento, California; [5]Visage Imaging, Inc., San Diego, California; and [6]Department of Radiology, Massachusetts General Hospital and Harvard Medical School, Boston, Massachusetts*

Artificial intelligence–generated content (AIGC) has shown remarkable performance in nuclear medicine imaging (NMI), offering cost-effective software solutions for tasks such as image enhancement, motion correction, and attenuation correction. However, these advancements come with the risk of hallucinations, generating realistic yet factually incorrect content. Hallucinations can misrepresent anatomic and functional information, compromising diagnostic accuracy and clinical trust. This paper presents a comprehensive perspective on hallucination-related challenges in AIGC for NMI, introducing the DREAM report, which covers recommendations for definition, representative examples, detection and evaluation metrics, and attributions and mitigation strategies. This position statement paper aims to initiate a common understanding for discussions and future research toward enhancing AIGC applications in NMI, thereby supporting their safe and effective deployment in clinical practice.

Artificial intelligence–generated content (AIGC) has demonstrated significant potential in nuclear medicine imaging (NMI) over the past decade, achieving state-of-the-art performance across various tasks, based on a range of quantification metrics. Key applications include PET or SPECT image enhancement such as denoising, deblurring, and partial-volume correction (*1*); quantitative accuracy improvements such as motion correction, scatter correction, and attenuation correction (AC) (*2*); and cross-modality image translation such as generating PET images from CT or MRI and vice versa (*3*). These artificial intelligence (AI)–driven solutions offer the potential to replace traditional hardware-dependent approaches with more cost-effective software alternatives while also potentially reducing radiation exposure, easing clinical workloads, and optimizing imaging workflows.

Despite these advancements, hallucinations pose significant challenges to AIGC in NMI applications. Hallucinations can lead to cascading errors, including misdiagnosis, mistreatment, unnecessary interventions, medication errors, and ethical or legal concerns (*4*). These risks highlight the urgent need for robust hallucination detection frameworks and mitigation strategies before AIGC can be safely deployed in clinical practice.

Although recent surveys (*5*,*6*) have explored hallucinations in natural language processing, the medical imaging community still lacks a domain-specific and systematic analysis of hallucinations. To bridge this gap, this paper presents a comprehensive perspective on hallucination-related challenges in AIGC for NMI. We introduce the DREAM report, which outlines key aspects including the definition of hallucinations, representative examples, detection and evaluation methods, and attributions and mitigation recommendations, as illustrated in Figure 1, with the main components summarized in Table 1.

## DEFINITION OF HALLUCINATIONS

The definition of hallucinations varies across publications and, in some cases, remains inconsistent or even contradictory. A precise and universally accepted definition has yet to be established (*7*). Although hallucinations were indistinguishable from general inaccuracies or errors in earlier studies (*8*,*9*), the term has gained renewed attention with the advent of large-scale and diffusion-based generative models (*10*,*11*). The enhanced generative capabilities introduce greater risks of fabricated content, prompting growing interest in establishing more rigorous and specific definitions of hallucinations.

In natural language processing, hallucinations are typically defined by their inconsistency with given targets (*12*). Factual hallucinations refer to AIGC that contradicts verifiable knowledge, whereas faithfulness hallucinations violate the instructions or source input. Other classifications include fact-conflicting, input-conflicting, and context-conflicting hallucinations (*13*). Additionally, Farquhar et al. (*14*) introduce a subset of hallucinations termed confabulations, referring to AIGC that are both incorrect and arbitrary. Here, arbitrary means that model outputs fluctuate unpredictably under identical inputs because of irrelevant factors such as random seed variations. This stochastic confabulation could be distinguished from systematic hallucinations in which AI claims are consistently incorrect, which may arise from flawed training data.

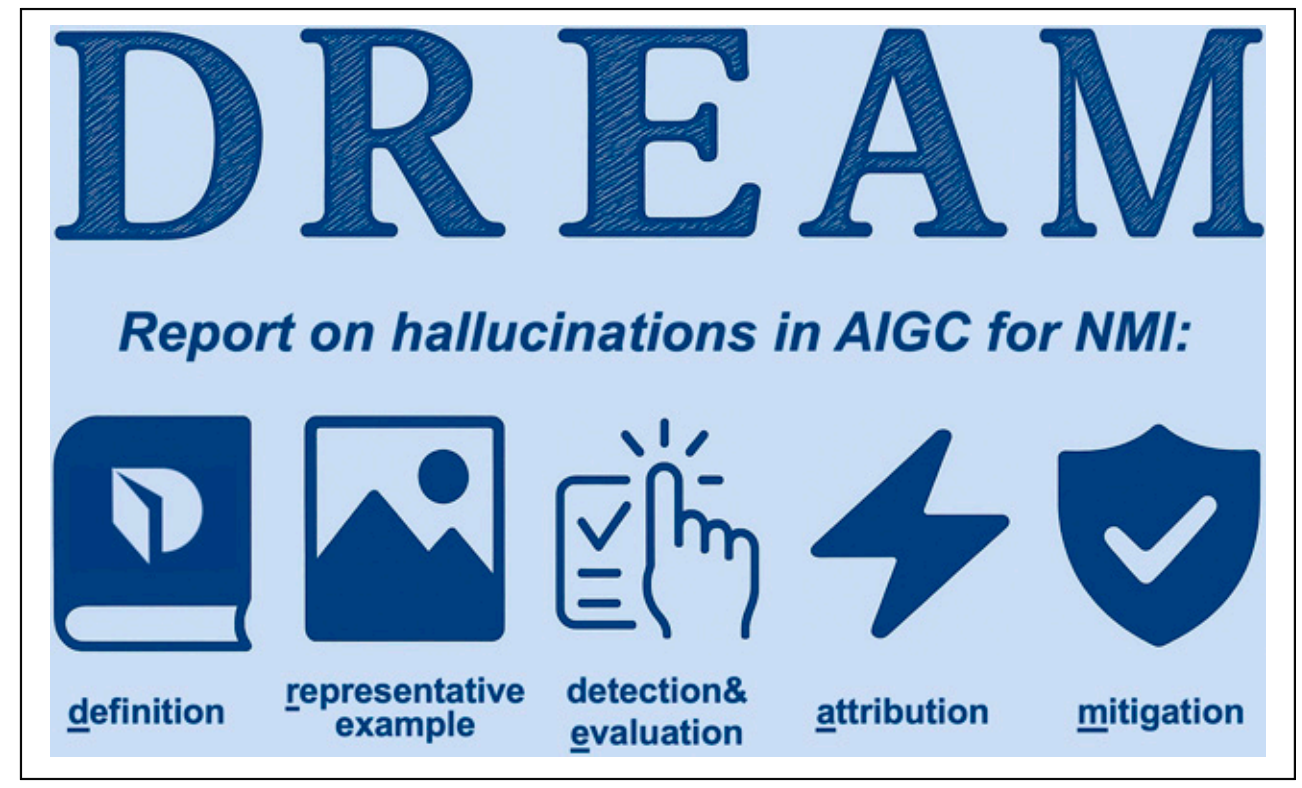

**FIGURE 1.** Organization of DREAM report.

In medical imaging, the definition of hallucinations remains similarly ambiguous. Some studies interpret hallucinations narrowly as the addition of nonexistent tissue components (*15*), whereas others encompass both the addition and removal of image structures (*16*,*17*), such as the omission of lesions (*18*,*19*). Certain researchers emphasize the deceptive and realistic-looking appearance of hallucinations (*20*), whereas others expand the scope to include implausible or dreamlike content (*21*,*22*). There is also disagreement about whether hallucinations are unique to AI. A study on tomographic image reconstruction (*23*) defines hallucinations as false structures in reconstructed images, regardless of origin. In contrast, others argue that hallucinations are unique to AI (*24*).

In this paper, we focus specifically on AIGC in NMI. Errors or artifacts introduced by traditional imaging workflows (supplemental material, section I; supplemental materials are available at http://jnm.snmjournals.org) (*25*) are considered out of the scope. An extensive literature review indicates that most AIGC applications in NMI operate as image-to-image translation tasks. In such settings, implausible large-scale errors, such as the addition of organs or major structures, are rarely observed. We argue that these dreamlike errors are better defined as delusions, borrowing from the psychologic lexicon. Instead, hallucinations in AIGC for NMI are typically subtle but deceptive, manifesting as added small abnormalities or realistic-looking lesions that do not exist in reality. Other plausible AI-induced errors, such as the omission of real lesions (falsely replacing abnormal regions with normal structures) or pure quantification bias (uniform intensity shifts without creating new structures), are better interpreted as illusions: misinterpreting something rather than fabricating. These forms of errors, although clinically significant, fall outside the scope of hallucinations in this paper.

Given this context, within the scope of this paper we recommend a narrow AI hallucination definition in NMI: AI-fabricated abnormalities or artifacts that appear visually realistic and highly plausible yet are factually false and deviate from anatomic or functional truth (or, in the case of NMI when ground truth images are unavailable, represent structures not supported by the measurement).

**NOTEWORTHY**

- Within this paper, hallucinations are defined as AI-fabricated abnormalities or artifacts that appear visually realistic and highly plausible yet are factually false and deviate from anatomic or functional truth.
- Hallucinations may occur across all AIGC applications in NMI.
- Recommended detection and evaluation methods include image-level comparisons, datasetwise statistical analysis, clinical task–based assessment (by human or model observers), and automated hallucination detectors trained on annotated benchmark datasets.
- Effective hallucination mitigation requires a comprehensive and multiperspective approach encompassing data quality, learning paradigms, and model design.
- Substantial adaptation and continued research are needed for robust detection, evaluation, and mitigation approaches tailored to AI hallucinations in NMI.

## REPRESENTATIVE EXAMPLES

In this section, we present visual examples of AIGC in representative NMI applications, illustrating scenarios in which hallucinations may arise during AI-driven processing.

### Image Enhancement

Numerous studies have explored AI-driven translation from low-count/high-noise to high-count/low-noise images (*26*–*28*), demonstrating impressive performance with outputs that are visually compelling and often nearly indistinguishable from high-quality reference scans. However, hallucinations can occasionally emerge, distorting underlying anatomic or functional information. Figure 2 presents examples of AI-driven SPECT and PET denoising, highlighting cases in which AIGC unintentionally alters critical imaging details.

### AC

AI-based AC techniques have been proposed to estimate AC images directly from non-AC images (*29*,*30*), eliminating the need for CT-based attenuation maps and their associated radiation dose and enabling use in dedicated imaging systems without integrated CT. Figure 3 presents examples of AI-driven AC in both PET and SPECT imaging. Although the AIGC appears visually accurate, closer comparison with reference CT AC images reveals hallucinations.

### Cross-Modality Translation

AI-driven cross-modality image translation synthesizes one imaging modality from another, enabling access to desired imaging information when only an alternative modality is available. For example, generating PET images from CT or MRI data (*31*,*32*) has been proposed to reduce the high costs and ionizing radiation exposure associated with PET. However, these potential benefits remain largely unproven in robust clinical trials, and the fundamental concept of inferring functional information from anatomic data (or vice versa) is inherently challenging and prone to significant pitfalls. In many conditions, functional abnormalities may appear before or independently of detectable structural changes, making such predictions unreliable. Cross-modality translation may be better suited as an auxiliary tool, such as for generating pseudo-CT scans from MRI for AC in PET/MRI workflows (*33*) or for augmenting training datasets with synthetic images to support AI model development.

Figure 4 presents examples of AI-driven cross-modality image translation, including PET–MRI (*31*), PET–CT (*3*), and PET–SPECT (*34*) conversions. Although these advancements highlight

**TABLE 1**
Summary of DREAM Report

| Category | Perspective | Attributions | Potential methods | Limitations/future work |
|---|---|---|---|---|
| Definition and examples* | | | | |
| Detection and evaluation | Image-level comparison | | Hallucination index; radiomics analysis | Consensus on annotation criteria; hallucination-specific metrics considering clinical relevance; interobserver and intraobserver variability studies; hallucination detector architecture; etc. |
| | Datasetwise statistical analyses | | Neural hallucination precursor; no-gold-standard evaluation | |
| | Clinical task assessment | | Performance on tasks such as lesion segmentation, disease classification; bounding boxes for localization; descriptive text annotations; and Likert scoring for severity and diagnostic quality | |
| | Automated hallucination detector | | Models trained on hallucination-annotated benchmark datasets | |
| Attributions and mitigation | Data | Domain shift | Guidelines on use specifying application ranges; improvement of data quality, quantity, and diversity; domain adaptation techniques; transfer learning; continuous data updates; and retrieval-augmented generation (RAG) | Data-efficient and generalizable approaches; fine-tuning of large foundation models using prior knowledge such as patient history; RAG techniques leveraging benchmark dataset as retrieval repositories; hallucination-aware mechanisms with hallucination detector providing feedback; etc. |
| | | Data nondeterminism | Optimization of data acquisition; data preprocessing, and cleaning | |
| | | Imperfect inputs or prompts | Optimization of input instructions; prompt engineering | |
| | Learning | Inherent probabilistic nature of deep learning | Feature/model averaging; user-guided interactive alignment; fast-checking system as defense layers | |
| | Model | Limited visual understanding or feature extraction | Using auxiliary perceptual information; using pathologic/structural constraints/priors | |

*AI-fabricated abnormalities or artifacts that appear visually realistic and highly plausible yet are factually false and deviate from anatomic or functional truth (or, in case of NMI in which ground truth images are unavailable, represent structures not supported by measurement). Examples of such hallucinations in AIGC for NMI are presented in "Representative Examples" section.

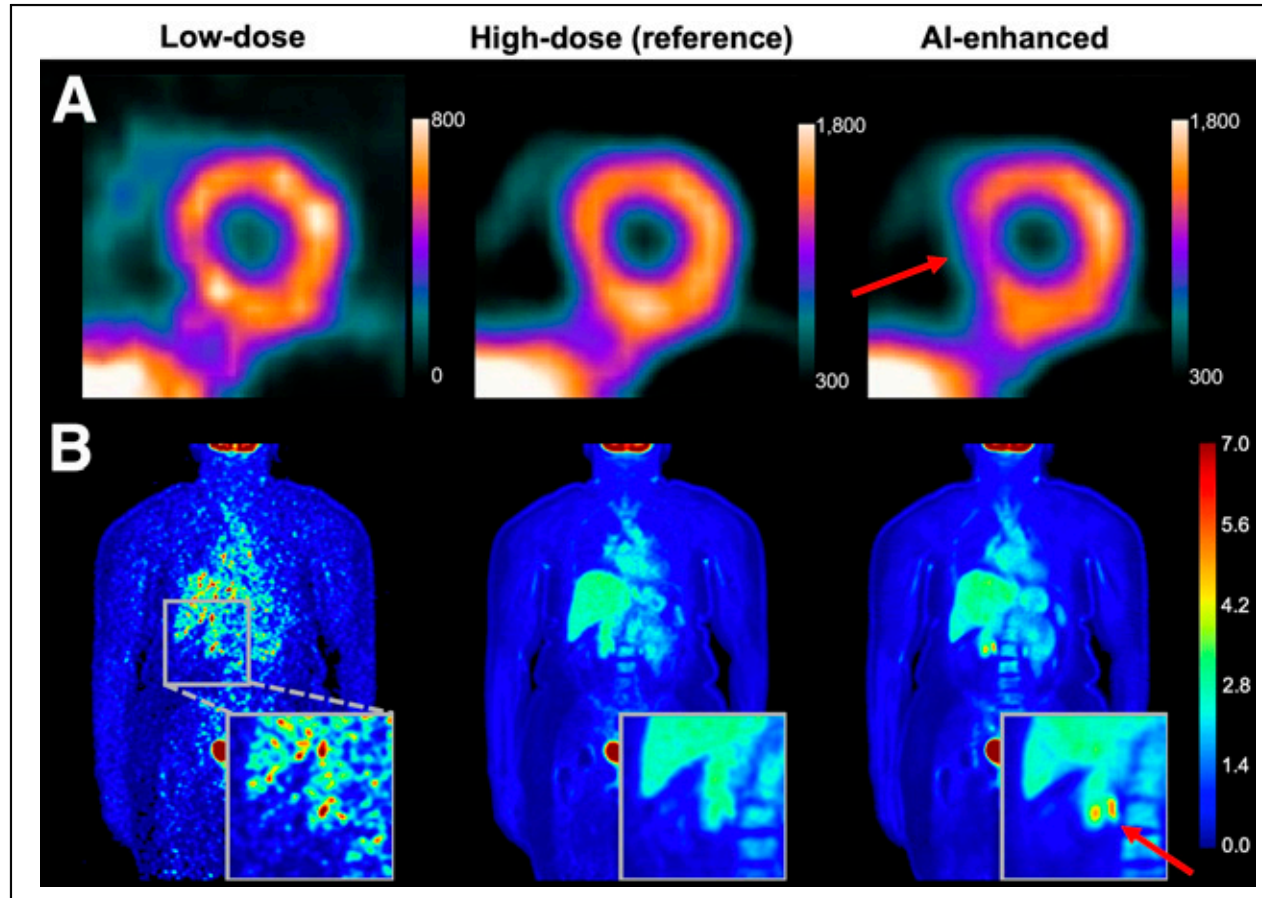

**FIGURE 2.** Examples of hallucinations in AI-driven image enhancement. (A) SPECT image denoising generated false-positive perfusion images (CC BY (*27*)). (B) PET image denoising generated false lesions, using method proposed by Dorjsembe et al. (*28*). Arrows indicate AI-introduced hallucinations.

the promising capabilities of AI in cross-modality imaging, they also underscore the risk of hallucinations.

## DETECTION AND EVALUATION METHODS

Although several commercial AI products, such as SubtlePET (Subtle Medical) and Precision DL (GE HealthCare), have already been cleared by the Food and Drug Administration for clinical use, the current Food and Drug Administration draft guidance on AI-enabled medical devices (January 2025) (*35*) does not explicitly reference the term *hallucination*. The guidance acknowledges that erroneous AI outputs can compromise device reliability and user trust and emphasizes a total-product-lifecycle approach spanning development, validation, and postmarket management. The guidance also recommends rigorous performance evaluation using metrics such as the area under the receiver operating characteristic curve and positive or negative likelihood ratios, among others. Although the Food and Drug Administration has jurisdiction over postdeployment monitoring, its oversight remains limited (e.g., adverse event reporting, required postmarket studies, and general device safety). Continuous performance monitoring in radiology practice is not mandated. To address this gap, the American College of Radiology has launched the Assess-AI initiative, with the long-term goal of influencing legislation to enable more robust postdeployment monitoring (*36*).

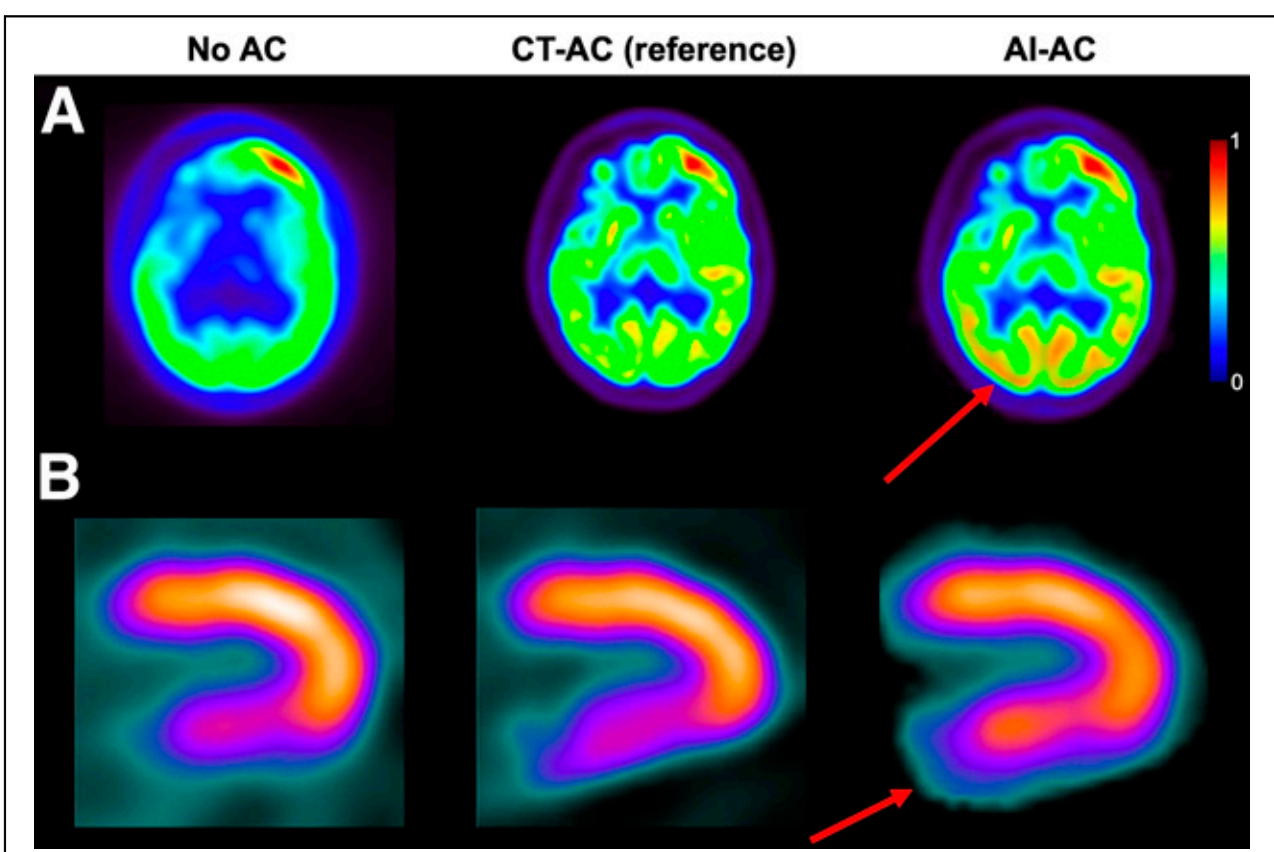

**FIGURE 3.** Examples of hallucinations in AI-driven CT-free AC for PET imaging, showing generated false abnormality in brain region (reprinted with permission of (*29*)) (A) and for SPECT imaging, showing generated false-negative perfusion (reprinted with permission of (*30*)) (B). Arrows indicate AI-induced hallucinations that caused overestimation of tracer concentration.

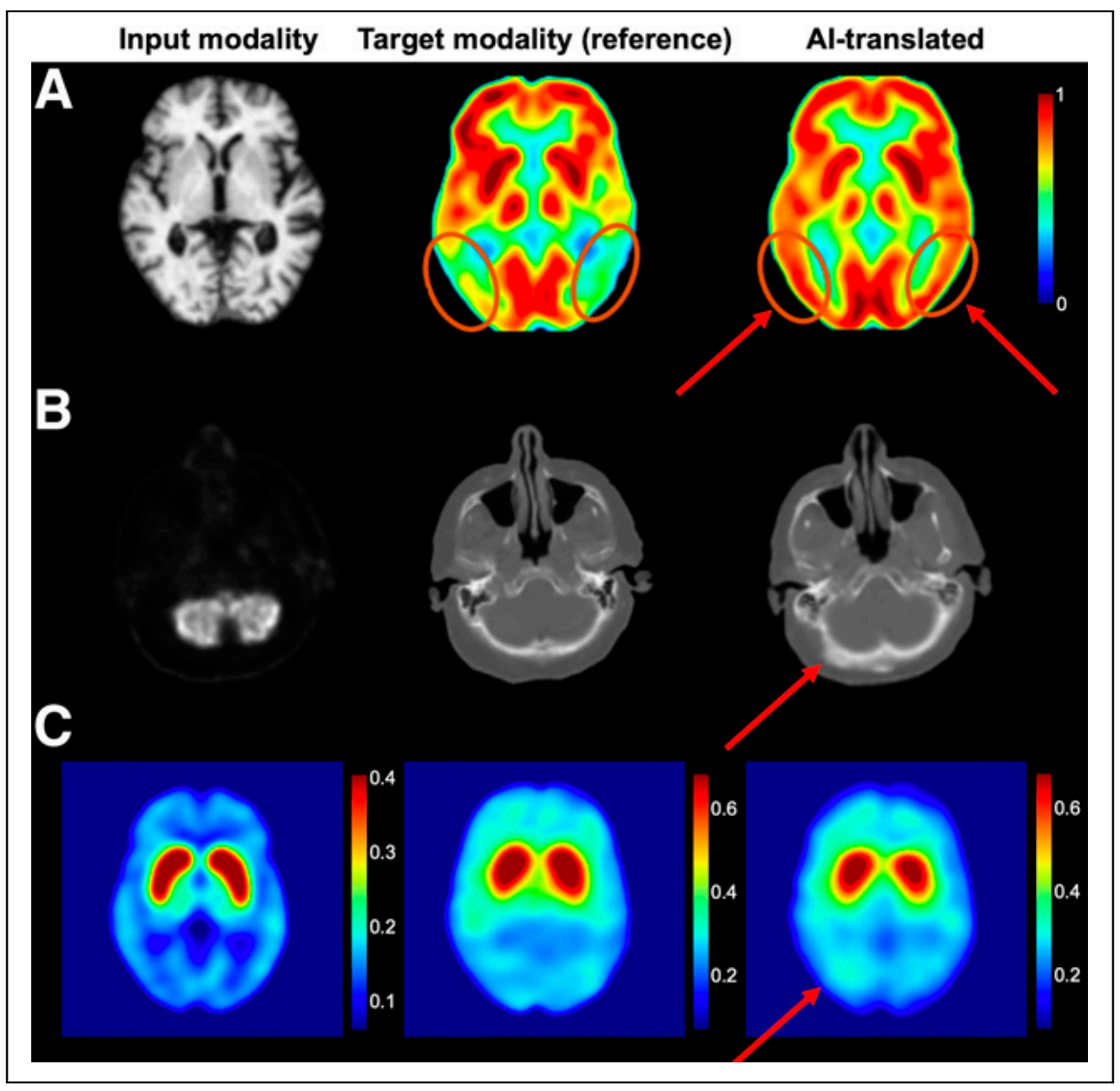

**FIGURE 4.** Examples of hallucinations in AI-driven cross-modality image translation. (A) PET/MRI translation (reprinted with permission of (*31*)). Although visually realistic, AI-generated PET image exhibits falsely increased glucose uptake in temporoparietal lobe, which could potentially lead to misdiagnosis of Alzheimer disease. (B) PET/CT translation (reprinted with permission of (*3*)). (C) PET/SPECT translation (CC BY (*34*)). Arrows highlight AI-induced hallucinations.

Within AI assessment and monitoring frameworks, hallucinations warrant dedicated and systematic attention. Instances identified during clinical use should be documented and reported through postmarket surveillance. In addition, establishing quantitative measures of hallucination could help define minimum acceptable thresholds for AI processing, such as the lowest-dose standards in AI-based denoising. Such thresholds would balance the extent of dose reduction with the risk of AI-induced hallucinations, ensuring that improved visual quality does not come at the cost of inaccurate representations.

To systematically assess hallucinations, we recommend adopting multifaceted metrics, which could draw on methodologies from related domains and be adapted to the specific context of NMI. Below, we outline several examples and preliminary ideas; however, further research is essential to develop accurate and widely accepted evaluation frameworks tailored to this field.

### Image-Based Metric

When paired reference images are available, the hallucination index has been proposed to detect AI-generated spurious features (*24*). This index is computed as the Hellinger distance between the distribution of AIGC and a so-called zero-hallucination reference. The latter is generated by adding adaptive white gaussian noise to the reference image, with the noise power calibrated to match the signal-to-noise ratio of the AIGC. However, the original formulation was tailored specifically for Fourier diffusion-based models, in which noise power is inferred via a diffusion bridge between the output and reference images. To extend the applicability of the hallucination index across diverse AI models, potential adaptations informed by insights from previous work (*37*,*38*) can be considered. Further details are provided in section II of the supplemental material.

Radiomics analysis has also been explored as an evaluation tool (*39*). Most AI models currently used in NMI prioritize visual image quality, often relying on loss functions such as mean squared error. Although such models produce outputs that appear visually of high quality, they do not necessarily improve data quality for downstream tasks and may introduce subtle errors and hallucinations. Radiomics-based evaluation detects this issue by selecting clinically relevant regions of interest and extracting quantitative features from both the AIGC and the corresponding reference images. Statistical comparisons between the 2 feature sets can reveal inconsistencies, with significant discrepancies potentially indicating the presence of hallucinations. However, not all discrepancies in radiomic features necessarily reflect hallucinations. Other types of errors that fall outside the hallucination definition adopted in this paper, such as lesion omission or pure quantification bias (as discussed in the "Definition" section), may also produce radiomic differences. Therefore, further research is needed to identify radiomic features that are specifically sensitive to hallucinations while minimizing confounding from other nonhallucinatory errors.

It is also worth noting that both the hallucination index and radiomics analysis primarily capture underlying statistical discrepancies between AIGC and the reference. If AI-generated artifacts alter only the visual appearance without affecting the statistical or diagnostic characteristics of the data, they may not be detected by these methods. This observation supports the distinction that although an artifact may change the appearance of an image, a hallucination alters the underlying data statistics. This interpretation aligns with our proposed hallucination definition, in which hallucinations are considered a subset of artifacts—specifically, those that are visually plausible but deviate from anatomic or functional truth. It underscores the conceptual understanding that not all artifacts are hallucinations but that hallucinations generally manifest as visually misleading artifacts.

#### Dataset-Based Metric

In scenarios in which paired reference images are unavailable, the neural hallucination precursor was introduced to quantify hallucinations from a feature-space perspective (*21*), under the assumption that false content arises from misrepresented features. The metric measures the k-nearest neighbor distance between the intermediate feature embeddings of the AIGC and a hallucination-free feature bank preconstructed from a calibration dataset sampled from the training set. However, the approach is inherently model-dependent, as the feature bank is defined and obtained by a specific model architecture. This dependence limits its applicability for comparing hallucination levels across different AI models, as each model may use distinct feature extraction mechanisms that shape the learned feature distributions differently.

To compare different models in the absence of reference images, the concept of no-gold-standard evaluation (*40*) may offer insights. Originally developed for assessing conventional quantitative imaging techniques, this method could be adapted for AIGC evaluation. It models a linear stochastic relationship between measured values and unknown true values, which are assumed to follow a 4-parameter beta distribution. Model parameters are estimated by maximizing the likelihood of the observed data, and the noise-to-slope ratio derived from these estimates is used to quantify the precision of each method (*40*). For AIGC evaluation, quantitative values could be defined as metrics such as the mean or maximum activity concentration within specific regions of interest in the AIGC. However, 2 key challenges arise. First, the assumed linearity between true and measured values may not hold for nonlinear generative models. Second, and as previously noted, this metric may capture general errors rather than hallucinations specifically. Adapting it to isolate hallucinations from other nonhallucinatory deviations will require substantial methodologic refinement.

#### Clinical Task–Specific Metric

Clinically relevant tools for hallucination detection and evaluation are essential for real-world deployment. One strategy assesses hallucinations indirectly through downstream segmentation or classification performance (*41*). Another relies on direct expert evaluation, in which medical professionals assess AIGC using disease-specific image features (*42*) or rate them on a 5-point Likert scale (section III of the supplemental material) (*43*). However, such scalar ratings alone are insufficient to capture the complexity of hallucinations. A more informative strategy could pair the Likert score with bounding box annotations that localize suspected hallucinations, accompanied by concise descriptive text (e.g., "a false, small lesionlike hot spot at the apex of the liver"). This offers greater granularity than scalar ratings while remaining more feasible for clinicians than full voxelwise segmentation. Nonetheless, these evaluations often require access to reference images; without them, even experienced readers may be misled by hallucinations. Furthermore, because it is impractical for physicians to review all generated cases, determining an adequate and representative sample size for hallucination evaluation remains a key challenge.

#### Automatic Hallucination Detector

Automatic hallucination detectors (*44*), trained on hallucination benchmark datasets (*4*), have recently been explored in large (vision) language models to reduce the burden of human evaluation (section IV of the supplemental material). However, to the best of our knowledge, no hallucination-annotated benchmark dataset currently exists for NMI applications. This underscores an urgent need for the research community to collaboratively develop such a resource, ideally through multiinstitutional efforts. A promising approach involves leveraging crowdsourcing platforms to collect a diverse set of AI-generated NMI images that exhibit hallucinations, along with expert annotations. These annotations could be guided by standardized criteria, such as bounding boxes to indicate hallucination locations, brief descriptive text, and Likert-scale ratings for severity, as discussed in the "Clinical Task–Specific Metric" section. Of course, the final annotation protocol will require further discussion and consensus to ensure consistency and clinical relevance. The dataset could be designed for continuous expansion, accommodating new contributions over time.

## ATTRIBUTIONS AND MITIGATION RECOMMENDATIONS

Most AIGC applications in NMI can be formulated as image-to-image estimation tasks, in which the objective is to learn a mapping function from a source domain $S$ to a target domain $T$, denoted as $G: S \rightarrow T$. Given a training dataset $\mathcal{D}$ with marginal distributions $P_S$ and $P_T$, the goal is to identify an optimal approximation: $\hat{G} = \text{argmin}_{\hat{G} \in H} \mathcal{L}(\hat{G}, \mathcal{D})$, where $\mathcal{L}$ is the loss function and $H$ the hypothesis space (*21*). Hallucinations arise when the learned mapping function $\hat{G}$ deviates from the true underlying mapping $G$. The mechanisms and attributions of hallucinations are complex and multifaceted, and mitigation strategies must be tailored to their specific causes, encompassing data quality, training paradigms, and model architecture (*15*,*41*).

### Data Perspective

**Domain Shift.** Domain shift, a mismatch between data distribution used for training and testing (i.e., test sample $s \notin P_S$), is widely recognized as a key contributor to hallucinations (*6*,*17*). Since generative AI models rely heavily on learned statistical priors, any deviation between training and testing distributions can result in unpredictable outputs, increasing the risk of hallucinations. For instance, overrepresentation of certain patterns in training data (e.g., lesions frequently occurring in the liver) may lead the model to erroneously hallucinate such features in test samples where they do not exist (*10*,*45*). Conversely, underrepresentation of certain pathologic scenarios may impair the model's performance on out-of-distribution samples, resulting in synthesized artifacts that do not correspond to actual medical conditions. A model trained primarily on healthy subjects, for instance, may hallucinate features when applied to rare diseases by extrapolating from incomplete or biased representations (*19*,*22*).

To mitigate hallucinations caused by domain shift, several strategies can be considered. First, guidelines on use should clearly define the intended scope and limitations of AI models, to prevent hallucinations caused by inappropriate or unintended applications. Second, improving the quality, quantity, and diversity of training data by including a wider range of scanners, imaging protocols, and patient populations can reduce hallucination risk. Figure 5A illustrates how richer and more comprehensive training datasets effectively decrease hallucinated artifacts. In a study by Zhou et al. (*46*), a federated learning network trained on datasets from 3 institutions outperformed a network trained on a single institution dataset. Third, when large-scale datasets are unavailable for training, domain adaptation techniques become useful. For instance, in a study by Liu et al. (*47*), an adversarial domain generalization method was used to handle PET denoising across arbitrary noise levels with training data limited to a narrow noise range. This method used a continuous discriminator to classify noise levels, thereby minimizing distribution shifts in latent feature representations across different noise domains. As shown in Figure 5B, the model incorporating this technique demonstrated reduced hallucinations compared with that trained without it. Fourth, transfer learning offers another effective solution by leveraging publicly pretrained models followed by fine-tuning on local data, striking a balance between generalization and specialization. For example, in another study by Liu et al. (*48*), a pretrained $^{18}$F-FDG PET denoising model was fine-tuned on only 3 $^{89}$Zr immuno-PET scans, enabling effective adaptation to a tracer-scarce target task. Under similar ideas, the concept of continuous dataset updating has been proposed, with models being regularly fine-tuned with newly acquired data to keep up with evolving clinical scenarios (*13*). However, such methods come with additional training costs and the potential risk of catastrophic forgetting. Fifth, retrieval-augmented generation provides an inference-time solution that improves output quality without retraining. For example, Shi et al. (*49*) reformulated complex medical questions into search-optimized synthetic queries, retrieving external knowledge from online databases to improve output quality. However, unlike language tasks, retrieval-augmented generation for NMI is currently limited because of the lack of well-structured, publicly available visual knowledge sources.

As discussed, each potential mitigation strategy presents advantages and limitations, particularly when applied to NMI. Substantial adaptation and continued research are needed to tailor these approaches to the unique challenges of this field.

**Data Nondeterminism.** The mapping function $G: S \rightarrow T$ is inherently nondeterministic because of aleatoric uncertainty in the dataset $\mathcal{D}$ (*21*). This nondeterminism arises from random variability in data acquisition processes, including measurement noise and stochastic fluctuations during image formation. The intrinsic ill-posedness of the estimation problem $\hat{G} = \text{argmin}_{\hat{G} \in H} \mathcal{L}(\hat{G}, \mathcal{D})$, given the dataset $\mathcal{D}$, results in one-to-many mappings for $G(s)$, where multiple plausible solutions may exist and many of them do not reflect the true observations. Consequently, this ambiguity can give rise to hallucinations.

To mitigate hallucinations caused by nondeterministic mappings, several strategies may be considered. First, optimizing data acquisition can help produce high-quality and consistent datasets, thereby establishing a more reliable foundation for model training. However, implementing this in practice is challenging, as it requires access to high-performance scanners and the ability to execute ultra-high-quality imaging protocols, particularly difficult for modalities such as SPECT and planar imaging. Second, applying rigorous data preprocessing, such as systematic data cleaning, can reduce inconsistencies and improve overall data fidelity. Nonetheless, substantial effort is required, and defining clear, objective criteria for determining whether the data meet quality standards remains a complex challenge.

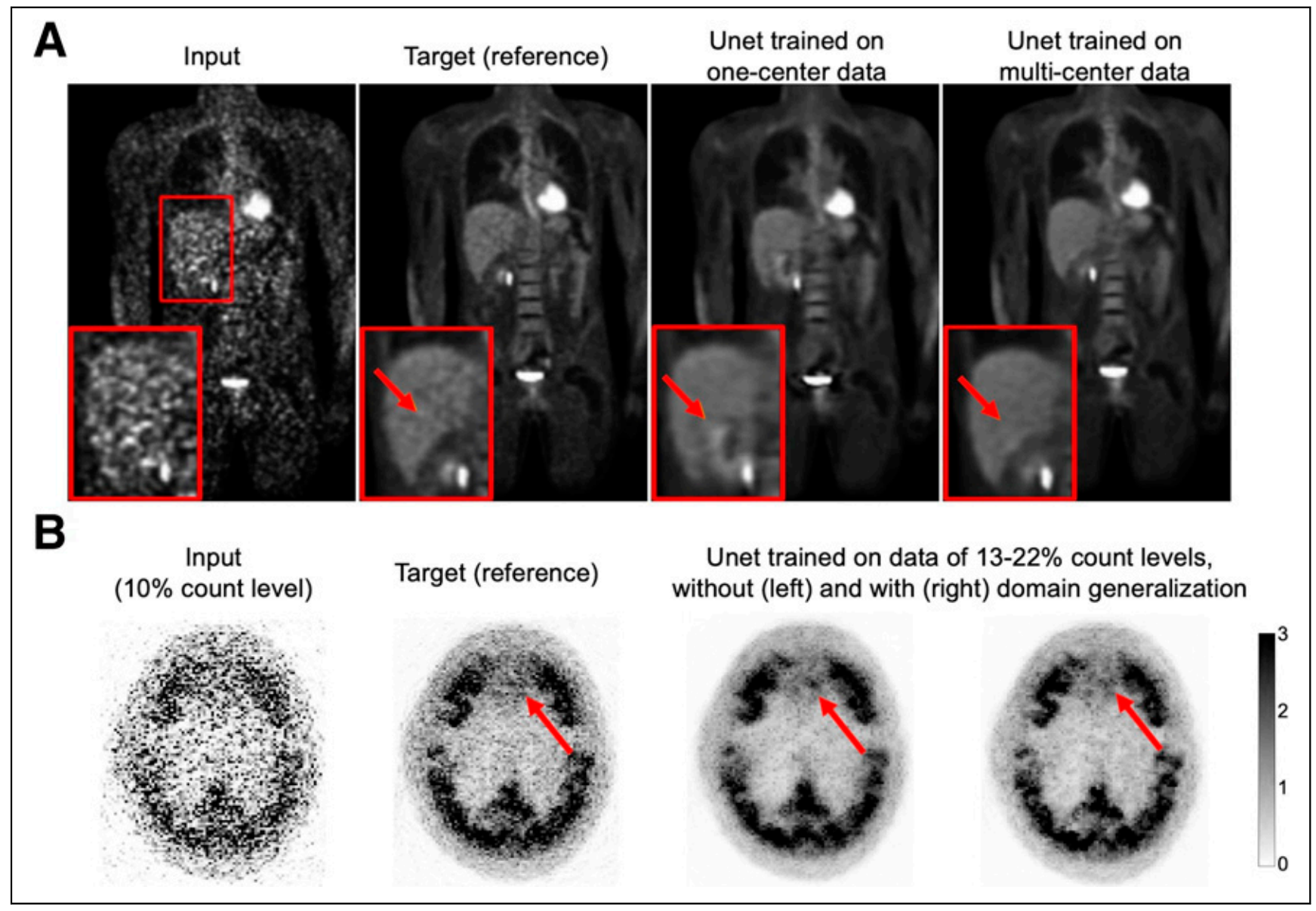

**FIGURE 5.** Examples of hallucination mitigation through use of richer and higher-quality training data for $^{18}$F-FDG whole-body PET denoising (reprinted with permission of (*46*)) (A) and domain generalization technique for $^{18}$F-MK-6240 brain PET denoising (adapted with permission of (*47*)) (B). Arrows indicate regions where hallucinations occur and their mitigation after applying corresponding proposed strategies.

In summary, although addressing aleatoric uncertainty at the data level holds promise for reducing hallucination risks, its practical implementation is often constrained by the high costs of hardware and the operational complexity of data acquisition.

**Input Perturbations or Imperfect Prompts.** Even in well-trained and high-performing AI models, hallucinations may still arise because of input perturbations or suboptimal prompts (*50*). Prompt engineering seeks to improve output accuracy by optimizing the structure and content of input instructions (*20*). Carefully formulated prompts that clearly define response boundaries and expectations help reduce ambiguity and guide the model toward more precise and reliable outputs (*13*). For instance, Yu et al. (*51*) introduced structured text prompts that explicitly specify organs and anatomic structures in the image, thereby enhancing the anatomic fidelity of denoised PET results. The accuracy of prompts plays a critical role in the model success. Similarly, Liu et al. (*52*) used dual prompts, one indicating noise count level and another providing a general denoising directive, to improve PET denoising across varying count levels.

### Learning Perspective

The inherent probabilistic nature of AI models makes hallucinations inevitable to some extent, analogous to the concept of epistemic uncertainty. AI models $\hat{G}$ rely on pattern recognition and statistical inference from training data, without a true understanding of meaning or facts. Consequently, hallucinations emerge as a fundamental limitation of data-driven learning systems (*21*). This inevitability arises from underspecification, in which many candidate solutions $\hat{G}$ within the Rashomon set $H^*$ can equally satisfy the training objective, that is, $\mathcal{L}_{\text{val}}(\hat{G},\mathcal{D}) < \tau, \forall \hat{G} \in H^*$, where $\mathcal{L}_{\text{val}}$ is the validation criteria and $\tau$ a predefined threshold. The Rashomon set $H^* \subset H$ comprises all models that achieve near-optimal performance within the possible space $H$ (*21*). Despite fitting the data well, these solutions may not align with the true underlying function, expressed as $\hat{G} \neq G$. Without a theoretic basis to prefer one solution over another, the randomly selected function $\hat{G}$ may deviate from ground truth, particularly in cases of small datasets or underconstrained generative frameworks such as unsupervised learning.

To mitigate this kind of hallucination, several techniques can be considered. First, ensemble model averaging or feature averaging, which aggregates outputs or latent features from multiple runs of models with similar architectures (i.e., multiple qualified $\hat{G}$ candidates), can reduce uncertainty and produce more stable results with fewer hallucinated artifacts. For example, in translating non-AC low-dose PET images into AC standard-dose PET images, Chen et al. (*53*) averaged outputs from three 2.5-dimension diffusion models across axial, sagittal, and coronal views, achieving better results than a single model run. Likewise, injecting random noise into inputs and averaging the resulting vision features across multiple runs have been shown to suppress spurious signals and improve reliability in medical image translation tasks (*54*). However, these averaging strategies incur high computational cost due to the need for multiple model runs. Second, user-guided interactive alignment may be especially valuable in the safety-critical context of NMI. This human-in-the-loop strategy uses iterative expert feedback to guide model learning toward better understanding of real-world facts (*55*). Practically, it involves incorporating human knowledge to interactively select the most plausible $\hat{G}$ solution from a pool of candidates. Although effective, this method is labor-intensive and subject to interobserver variability. Third, to alleviate human workload, automated fast-checking systems have been developed to simulate expert feedback and interactions (*13*,*56*). These systems leverage predefined rules, statistical heuristics, or learned hallucination detectors (as discussed in the “Automatic Hallucination Detector” section), to flag potentially erroneous content in model outputs. Serving as an auxiliary verification layer or adversarial critic, these systems enhance the reliability and interpretability of AI-generated outputs. Nevertheless, the effectiveness of this approach depends on the accuracy and robustness of the checking system itself.

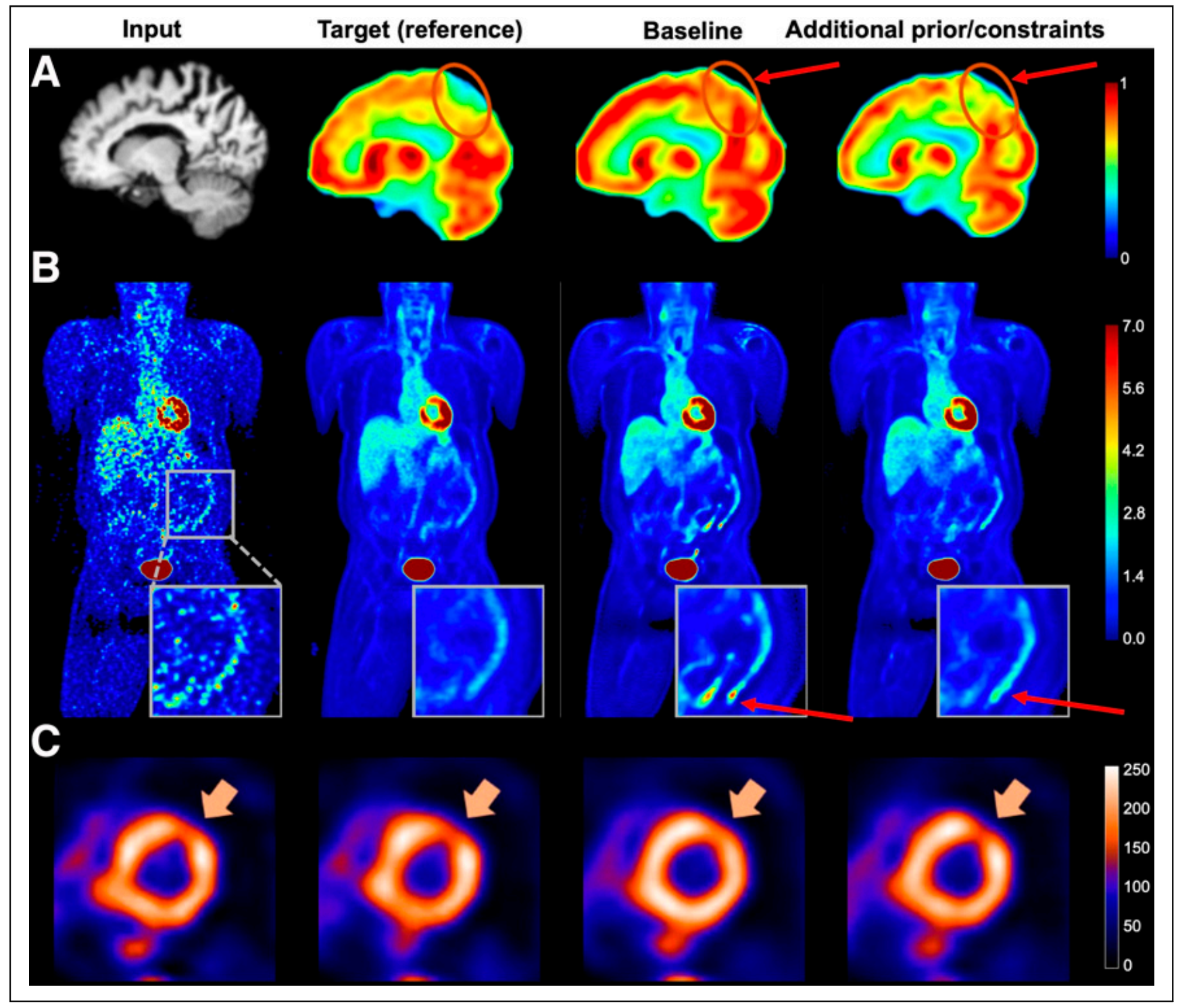

**FIGURE 6.** Examples of hallucination mitigation using additional constraints or priors. (A) MRI-to-PET translation incorporating multimodal and clinical prior information (reprinted with permission of (*31*)). (B) PET denoising using anatomic and metabolic priors to regularize generation (*57*). (C) SPECT denoising guided by task-specific loss function informed by perfusion defect detection (CC BY (*58*)). Circles and arrows indicate regions where hallucinations occur and their mitigation after applying corresponding proposed strategies.

### Model Perspective

Another key contributor to hallucinations may be the AI model's limited capacity for visual understanding and feature learning, directly impacting the reliability of its outputs.

Improving the perceptual capability of vision encoders can be achieved through more context-appropriate architectural designs and the integration of additional perceptual information, such as semantic maps or multimodality representations. For example, a pathology-aware translation model was proposed for PET/MRI image translation (*31*). It used adaptive group normalization layers to integrate multimodal conditions, including demographic information, cognitive scores, and Alzheimer disease biomarkers. These fused multimodal priors enhanced the preservation of pathologic features in the generated PET images, compared with the baseline model without such conditioning, as illustrated in Figure 6A. In addition, incorporating strong anatomic and functional constraints, through either auxiliary encoders or specialized loss functions, has shown promise in reducing hallucinations by guiding more robust feature extraction. For example, in a study by Xia et al. (*57*), an anatomically and metabolically informed diffusion model was introduced for PET denoising. This model incorporated lesion and organ segmentation maps as auxiliary constraints to regularize the denoising process, improving structural fidelity in generated PET images (Fig. 6B). Similarly, in a study by Rahman et al. (*58*), a task-specific loss term was added to a baseline SPECT denoising model, incorporating performance on perfusion defect detection as an auxiliary supervision signal. This addition helped suppress hallucinations in the denoised outputs, as shown in Figure 6C. Although effective, these approaches often incur additional computational cost due to the integration of complex priors and regularization mechanisms.

## FUTURE WORK

Despite the recommendations and insights outlined in this paper, continued research and substantial work are needed. The strategies presented here may encounter limitations when applied to specific NMI scenarios. Moreover, if AI is intended to enhance clinical workflow efficiency, any associated detection or mitigation strategies must be designed to minimize time burden during deployment. Additional discussion on future directions and potential implementations is provided in section V of the supplemental material.

## CONCLUSION

Hallucinations in AIGC for NMI remain a critical challenge. In this paper, we introduce the DREAM report, which offers a comprehensive perspective on hallucinations in AIGC for NMI. These hallucinations may arise from biased or nondeterministic data, the intrinsic probabilistic nature of deep learning, or limited visual feature understanding by models. Effective detection and evaluation require multifaceted frameworks, incorporating image-based, dataset-based, and clinical task–based metrics, as well as the development of automated detectors trained on hallucination-annotated datasets. Mitigation strategies must be tailored to the specific causes of hallucinations and should involve enhancements in data quality, learning methodologies, and model architectures. This DREAM report serves as a starting point for discussions in the field, highlighting the need for continued and extensive research.

## DISCLOSURE

This work was supported by the National Institutes of Health (NIH) under grants R01CA275188 and P41EB022544. No other potential conflict of interest relevant to this article was reported.

## ACKNOWLEDGMENT

This article was published as a preprint (https://arxiv.org/abs/2506.13995).

## Supplemental Data

### I. Discussions on artifacts from non-AI processes

While this DREAM report focuses on AI hallucinations, it is important to recognize that analogous concepts and concerns have long existed in NMI prior to the advent of AI. Traditional image acquisition, reconstruction, and processing pipelines are inherently susceptible to various artifacts, some of which can be visually misleading and described as 'hallucinations' in a broader or historical context (23). AI hallucinations and traditional artifacts may share certain visual similarities, but they differ fundamentally in origin, mechanisms, and interpretability.

Conventional tomographic reconstruction algorithms, such as filtered back projection (FBP) and iterative methods (e.g., OSEM), are known to produce artifacts, including streak artifacts in FBP; smoothing effect in iterative reconstruction due to regularization; convergence bias, where over- or under-iteration alters lesion contrast; and motion artifacts from patient movement, respiration, or cardiac motion during acquisition. While these artifacts can impact image interpretation, they tend to follow predictable patterns and are generally recognizable to experienced clinicians. In contrast, AI hallucinations are often more variable and unpredictable, making them harder to detect and potentially more dangerous in clinical decision-making.

Artifacts can also arise from non-AI post-processing steps. For example, (25) demonstrated that summation of gated interval reconstructions can produce false-negative myocardial filling of a defect in SPECT images. Such artifacts can closely mimic the visual appearance of AI hallucinations.

Overall, artifacts from conventional imaging are typically interpretable; their causes, such as respiratory motion or inappropriate filter settings, could usually be identified. By contrast, hallucinations generated by AI models are generally non-deterministic, opaque, and difficult to trace to specific inputs or processing steps, making quality control and hallucination analysis in AI pipelines inherently more challenging.

### II. Adapting the hallucination index computation

Taking whole-body PET denoising as an example, the signal-to-noise ratio (SNR) estimation methodology proposed in prior studies (37)(38) can be adopted to align the SNR of a high-count reference image with that of an AIGC output. This alignment facilitates the construction of a comparable 'zero-hallucination reference' image, as originally defined in (24). Specifically, SNR is calculated as the ratio of the mean to the standard deviation of voxel intensities within a uniform spherical region of interest, typically 3 cm in diameter and placed in biologically stable organs such as the liver. Adaptive white Gaussian noise is added to the image with higher SNR, such that the SNRs of the AIGC and the reference are matched. The Hellinger distance is then computed between distributions of the AIGC and the constructed reference. By computing a baseline distance between the low-count input and the high-count reference in the same manner, the ratio of the AIGC-to-reference distance to this baseline can be used as the final hallucination index. A visual comparison is presented in Supplemental Figure 1, where hallucination indices are computed for denoised outputs from two different models. The results show that a lower hallucination index can be associated with fewer hallucinated artifacts, supporting the metric's potential utility as a quantitative tool for assessing AI-induced hallucinations in NMI.

Even so, several limitations of the above hallucination index formulation must be acknowledged. **First**, while the index can be used for relative comparisons between different generated results, where a lower value may suggest fewer hallucinations, the interpretation of its absolute value remains ambiguous. Specifically, it is unclear what value range constitutes severe hallucinations or what threshold might be deemed clinically unacceptable. **Second**, the construction of a 'zero-hallucination reference' by adding white Gaussian noise cannot accurately account for complex characteristics in PET imaging, which may be better modeled by Poisson or mixed Poisson-Gaussian distributions. Moreover, the current SNR estimation reflects only local noise behavior and may fail to reflect variability outside the selected region of interest. **Third**, the current formulation does not consider anatomical context or clinical relevance of hallucinations. For instance, hallucinated lesions or hot spots in critical organs such as the liver or lungs could have a much greater diagnostic impact in $^{18}$F-FDG oncology imaging than similar artifacts in less critical regions like the bowel. A potential improvement could be incorporating organ segmentation maps into the index calculation, enabling spatially weighted assessments that emphasize hallucinations in diagnostically important areas.

In summary, substantial research is still needed to develop a more representative and broadly applicable hallucination index in NMI.

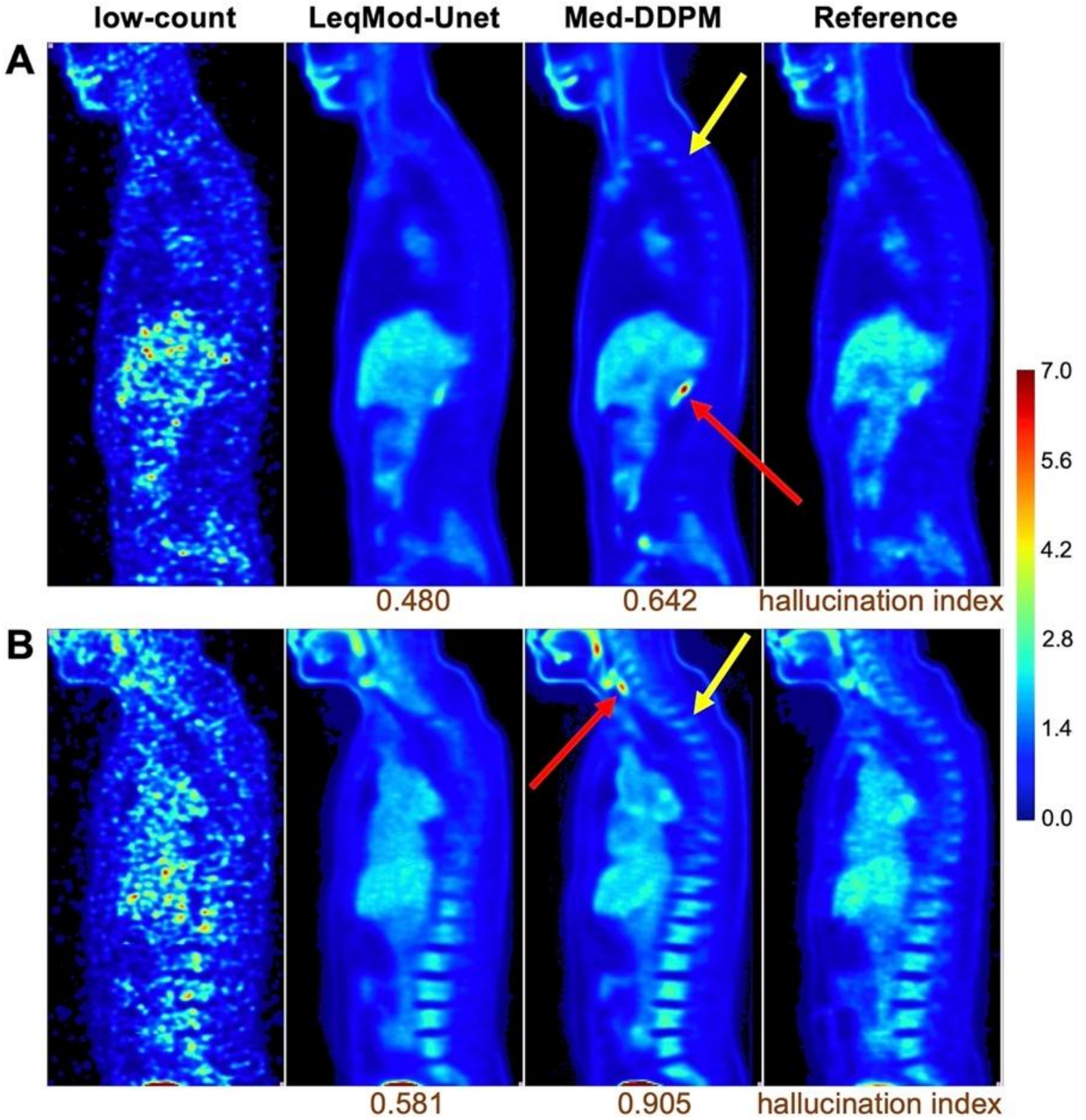

Supplemental Figure 1: Examples of hallucination index computation in PET image denoising. From left to right: input low- count images, denoised results using LeqMod-Unet (26) and Med-DDPM (28), and the high-count reference image. The hallucination index between each AI-generated image and the reference is displayed below the im- age, enabling a quantitative comparison. Notably, the diffusion-based Med-DDPM produces results with clearer anatomical structures and higher visual quality (highlighted by yellow arrows), but it also demonstrates a greater tendency to hallucinate false content (indicated by red arrows).

## III. Examples of clinical task-specific metrics

Some studies evaluate hallucinations though downstream segmentation or classification tasks (39)(41). In these approaches, task-specific observers, either expert radiologists or pre-trained disease classifiers/segmenters, analyze both the AIGC and corresponding reference images. Their decisions are then compared to identify potential hallucinations. For example, in a denoising evaluation study (39), expert readers performed segmentation on both denoised and high-dose images. The Dice coefficient was calculated to quantify segmentation agreement, serving as a hallucination indicator. If a hallucinated lesion appears in the denoised image, the resulting segmentation mismatch would yield a reduced Dice score. This strategy, however, has notable limitations. Human reading is time-consuming and often infeasible for large-scale datasets, while automated observers may introduce their own errors, potentially confounding accurate hallucination detection.

**5-point Likert scale for hallucination severity:**

5 (Excellent): No detectable hallucinations; high diagnostic quality.

4 (Good): Minor hallucinations; acceptable for clinical interpretation.

3 (Average): Noticeable hallucinations but still interpretable (e.g., added lesions/artifacts in diagnostically irrelevant regions).

2 (Below Average): Significant hallucinations that impair interpretation.

1 (Poor): Severe hallucinations; unsuitable for clinical use.

## IV. Examples of automatic hallucination detectors in large (vision) language models

Automatic hallucination detectors have been developed by training on benchmark hallucination datasets. For example, the Med-HallMark dataset (4) comprises paired input images and model-generated text, each annotated by human experts for hallucinations. Annotations categorize outputs into six levels: catastrophic, critical, attribute, prompt-induced, minor, and correct statements, each assigned a weight within the proposed MediHall score: 0.0, 0.2, 0.4, 0.6, 0.8, and 1.0, respectively. Higher MediHall scores indicate fewer hallucinations. Using this annotated dataset, a MediHall detector was trained to automatically classify hallucination severity and output corresponding MediHall scores for medical text produced by large language models.

In the vision domain, frameworks such as AQuA (16) have been developed to assess hallucinations in AI-driven image translation tasks. AQuA introduces various morphological hallucinations and error types into translated images, which are then manually annotated and labeled as acceptable or unacceptable. A dedicated model, AQuA-Net, was subsequently trained on this dataset to automate hallucination detection and assess the reliability of generated results.

## V. Discussions on future work

A foundational next step toward standardizing the evaluation and detection of AI hallucinations in NMI may be the development of a publicly available, large-scale benchmark dataset tailored for AIGC in NMI. This dataset should include a wide range of AI-generated NMI images, ideally paired with corresponding reference images when available, while also accommodating unpaired cases. Most importantly, each image should be accompanied by structured hallucination annotations. The diversity of generative models across various NMI applications provides a solid foundation for curating such a resource, which could support both benchmarking and the development of automated hallucination detectors. Several critical considerations must be addressed in constructing such a benchmark. **First**, while a comprehensive benchmark spanning multiple NMI modalities would be ideal, significant differences in tracer characteristics, spatial resolutions, and diagnostic contexts make cross-modality standardization highly challenging. A more feasible approach is to initially develop modality-specific benchmarks, with potential for gradual integration over time. **Second**, establishing a consensus annotation protocol is essential. As discussed in Sections IV.3 and IV.4, annotations could include bounding boxes for localization, short descriptive texts summarizing hallucination characteristics, and Likert-scale scores to reflect severity and diagnostic impact. When feasible, richer annotations such as lesion segmentations or disease classifications, especially for those likely to be influenced by hallucinations, could be incorporated. To reduce annotation burden, an interactive annotation loop can be employed, combining expert-reviewed samples, trained hallucination detectors, and visualization tools, to enable scalable annotation with human oversight. **Third**, the benchmark should support computational evaluation metrics in addition to human annotations. Existing methods such as the hallucination index and no-gold-standard evaluation (Sections IV.1 and IV.2) provide initial references, but they do not adequately isolate hallucinations from general errors. New metrics should be developed based on features or statistical patterns that are specifically sensitive to hallucinations, while remaining robust against non-hallucinatory errors. Importantly, such metrics should reflect anatomical and clinical relevance, potentially by assigning greater weights to hallucinations in diagnostically critical regions (e.g., liver, lungs) compared to those in less critical areas (e.g., bowel). **Fourth**, hallucination evaluation in cases without reference images remains particularly challenging. In these scenarios, both expert readers and automatic detectors may be misled by highly realistic outputs. Here, inter- and intra-observer variability could serve as an auxiliary signal, capturing uncertainty or disagreement among clinicians in interpreting potential hallucinations. **Fifth**, the design of hallucination detector architecture warrants further investigation. Ideally, models should handle AIGC inputs with and without reference, and support multi-headed outputs to jointly predict hallucination presence, spatial location, descriptive attributes, severity, and task-specific risks. Fine-tuning existing large vision language models may be a promising direction; however, most large vision language models are optimized for natural 2D images, so adapting them to volumetric NMI remains a key technical hurdle.

On the mitigation side, while access to high-quality and large-scale training data remains the ideal approach to mitigating hallucinations, it is often impractical in NMI due to limited data availability and inherent variability in real-world acquisitions (as noted in Section V.1.2). Therefore, future research should prioritize data-efficient and generalizable solutions. Several directions merit further exploration. For instance, transfer learning from high-resource modalities or domain adaptation techniques may enhance generalizability when NMI data are limited. Alternatively, fine-tuning large vision-language foundation models pretrained in other domains, informed by prior knowledge such as patient history, offers a compelling strategy. In addition, retrieval-augmented generation techniques could be adapted for vision tasks, using the curated hallucination benchmark dataset as a retrieval repository to guide outputs toward more accurate. Another promising direction involves integrating automated hallucination detectors into the training process. For example, hallucination-aware loss functions, such as adversarial terms informed by hallucination detector feedback, could be employed to penalize hallucination-prone generations. These approaches may help proactively suppress hallucinations and enhance the reliability of AIGC in NMI.

Ultimately, embedding hallucination detection and correction into clinical pipelines remains a critical yet largely unexplored frontier. This will require the development of real-time, computationally efficient hallucination screening tools and hallucination-aware systems that can function seamlessly in clinical workflows. To ensure practical adoption, these systems must operate without introducing significant time burden, thereby enabling AIGC to enhance, rather than compromise, diagnostic safety and effectiveness in NMI.